\documentstyle[psfig,aps,prl,multicol]{revtex}
\begin{document}

\title{Quantum cloning and the capacity of the Pauli channel}

\author{Nicolas J. Cerf$^{1,2,3}$}
\address{$^1$W. K. Kellogg Radiation Laboratory,
California Institute of Technology, Pasadena, California 91125\\
$^2$Information and Computing Technologies Research Section,
Jet Propulsion Laboratory, Pasadena, California 91109\\
$^3$Center for Nonlinear Phenomena and Complex Systems,
Universit\'e Libre de Bruxelles,
1050 Bruxelles, Belgium}

\date{March 1998}

\draft
\maketitle
\begin{abstract}
A family of quantum cloning machines is introduced
that produce two approximate copies from a single
quantum bit, while the overall input-to-output operation 
for each copy is a Pauli channel.
A no-cloning inequality is derived, describing the balance 
between the quality of the two copies. 
This also provides an upper bound on the quantum capacity
of the Pauli channel with probabilities $p_x$, $p_y$ and $p_z$.
The capacity is shown to be vanishing if
$(\sqrt{p_x},\sqrt{p_y},\sqrt{p_z})$ lies outside an ellipsoid
whose pole coincides with the depolarizing channel that underlies 
the universal cloning machine.
\end{abstract}
\pacs{PACS numbers: 03.67.Hk, 03.65.Bz, 89.70.+c
      \hfill KRL preprint MAP-220}


A remarkable consequence of the linearity of quantum mechanics is
that an unknown quantum state {\em cannot} be duplicated, 
as recognized after the seminal papers by Dieks~\cite{bib_dieks}, and
Wootters and Zurek~\cite{bib_wz}.
This so-called {\em no-cloning} theorem implies, for instance, that
there exists no physical process that can produce
{\em perfect} copies of a quantum bit (qubit) that is initially 
in an arbitrary state $|\psi\rangle=\alpha |0\rangle + \beta |1\rangle$.
Recently, it has been shown by Buzek and Hillery~\cite{bib_bh} that
it is nevertheless possible to construct a cloning machine
that yields two {\em approximate} copies of an input qubit.
Specifically, a universal cloning machine (UCM) can be defined
that creates two copies characterized each by the same density operator $\rho$
from a single qubit in state $|\psi\rangle$, the fidelity of cloning
being $f\equiv\langle\psi|\rho|\psi\rangle=5/6$. The UCM 
was later proved to be optimal by Bruss et al.~\cite{bib_bruss},
and Gisin and Massar~\cite{bib_gisinmassar}.
This cloning machine is {\em universal}
in the sense that the copies are state-independent (both output qubits
emerge from a depolarizing channel of probability 1/4, that is,
the Bloch vector characterizing the input qubit is shrunk by a factor
2/3 regardless its orientation).
A great deal of effort has been devoted recently to quantum cloners
because of their use in the context of quantum communication and
cryptography (see, e.g.,~\cite{bib_bruss,bib_gisinhuttner}).
For example, an interesting application of the UCM is that
it can be used to establish an upper bound on the quantum capacity $C$
of a depolarizing channel, namely $C=0$ at $p=1/4$~\cite{bib_bruss}.
\par

In this paper, I introduce a family of asymmetric cloning machines
that produce two (not necessarily identical) output qubits,
each emerging from a Pauli channel.
This family of cloners, which I call {\em Pauli cloning machines} (PCM),
relies on a parametrization of 4-qubit wave functions
for which all qubit pairs are in a mixture of Bell states. 
Using these PCMs, I derive a {\em no-cloning inequality}
governing the tradeoff between the quality of both copies
of a single qubit imposed by quantum mechanics. It is, by construction,
a tight inequality for qubits which is saturated using a PCM. 
I then consider a subclass of {\em symmetric} PCMs in order to
express an upper limit on the quantum capacity of a Pauli channel,
generalizing the considerations of Ref.~\cite{bib_bruss}
for a depolarizing channel.
\par

\begin{figure}
\caption{Pauli cloning machine of input $x$ and outputs $y_1$ and
$y_2$. The third output $y_3$ refers to an idle qubit (or 
the cloning machine). The three outputs emerge in general
from distinct Pauli channels.}
\vskip 0.25cm
\centerline{\psfig{file=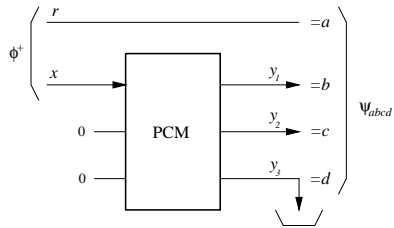,width=2.0in,angle=-90}}
\label{fig_cloner}
\vskip -0.25cm
\end{figure}

When processed by a Pauli channel, a qubit 
is rotated by one of the Pauli matrices or remains unchanged:
it undergoes a phase-flip ($\sigma_z$), a bit-flip ($\sigma_x$), 
or their combination ($\sigma_y$) with
respective probabilities $p_z$, $p_x$, and $p_y$. (A depolarizing
channel corresponds to the special case $p_x=p_y=p_z$.)
If the input qubit $x$ of the Pauli channel is initially
in a fully entangled state
with a {\em reference} qubit $r$, say in the Bell state $|\Phi^+\rangle$,
then the joint state of $r$ and the output $y$  
is a mixture of the four Bell states
$|\Phi^{\pm}\rangle = 2^{-1/2} (|00\rangle \pm |11\rangle)$ and
$|\Psi^{\pm}\rangle = 2^{-1/2} (|01\rangle \pm |10\rangle)$,
\begin{equation}   \label{eq_mixture-Bell}
\rho_{ry} = 
 (1-p)\, |\Phi^+\rangle\langle \Phi^+|
+p_z \, |\Phi^-\rangle\langle \Phi^-|
+p_x \, |\Psi^+\rangle\langle \Psi^+|
+p_y \, |\Psi^-\rangle\langle \Psi^-|  \; ,
\end{equation}
with $p=p_x+p_y+p_z$.
For an asymmetric cloning machine whose two outputs, $y_1$ and $y_2$,
emerge from (distinct) Pauli channels, the density operators $\rho_{ry_1}$
and $\rho_{ry_2}$ must then be mixtures of Bell states.
Focusing on the first output $y_1$, we see that
a 4-dimensional Hilbert space is necessary in general
to purify $\rho_{ry_1}$ since we need to accommodate its four
(nonzero) eigenvalues.
The 2-dimensional space of second output qubit $y_2$ is thus 
insufficient for this purpose,
so that we must introduce an additional qubit $y_3$,
which may be viewed as an ancilla or the 
cloning machine itself~\cite{fn0}. Thus, we are led to consider
a 4-qubit system in order to describe the PCM,
as represented in Fig.~\ref{fig_cloner}.
Before cloning, the qubits $r$ and $x$ are in the entangled state 
$|\Phi^+\rangle$, the two auxiliary qubits being in a prescribed
state, e.g., $|0\rangle$.
After cloning, the four qubits $r$, $y_1$, $y_2$, and $y_3$
are in a pure state for which $\rho_{ry_1}$ and $\rho_{ry_2}$ are
mixtures of Bell states. As we shall see below, $y_3$ can be viewed as
a third output (or ``idle'' qubit) which also emerges
from a Pauli channel.
\par

Thus, instead of specifying a PCM by a particular
unitary operation acting on the state 
$|\psi\rangle=\alpha |0\rangle + \beta |1\rangle$
of the input qubit $x$ (together with the two auxiliary qubits
in a fixed state $|0\rangle$), it is more convenient
to characterize it by the
wave function $|\psi\rangle_{r y_1 y_2 y_3}$ underlying the
entanglement of the three outputs with $r$.
So, the question here is to find in general 
the 4-qubit wave functions $|\psi\rangle_{abcd}$
that satisfy the requirement that the state of every pair of two qubits is
a mixture of the four Bell states. 
Making use of the Schmidt decomposition 
for the bipartite partition $ab$ vs $cd$, it is clear that
$|\psi\rangle_{abcd}$ can be written as a superposition
of {\em double Bell} states~\cite{fn1}
\begin{equation}  \label{eq_psi}
|\psi\rangle_{abcd}= 
\left\{ v \, |\Phi^+\rangle |\Phi^+\rangle 
+z \, |\Phi^-\rangle |\Phi^-\rangle 
+x \, |\Psi^+\rangle |\Psi^+\rangle 
+y \, |\Psi^-\rangle |\Psi^-\rangle  \right\}_{ab;cd}  \; ,
\end{equation}
where $x$, $y$, $z$, and $v$ are parameters
($|x|^2+|y|^2+|z|^2+|v|^2=1$).
The above requirement is then satisfied
for the qubit pairs $ab$ and $cd$, that is,
$\rho_{ab}=\rho_{cd}$ is of the form of Eq.~(\ref{eq_mixture-Bell})
with $p_x=|x|^2$, $p_y=|y|^2$, $p_z=|z|^2$, and $1-p=|v|^2$.
It is worth noting that these double Bell states for the partition
$ab$ vs $cd$ transform into superpositions of double Bell states 
for the two other possible partitions ($ac$ vs $bd$, $ad$ vs $bc$), e.g.,
\begin{equation}
|\Phi^+\rangle_{ab} |\Phi^+\rangle_{cd} =
{1\over 2} \left\{ |\Phi^+\rangle |\Phi^+\rangle 
+|\Phi^-\rangle |\Phi^-\rangle 
+|\Psi^+\rangle |\Psi^+\rangle 
+|\Psi^-\rangle |\Psi^-\rangle  \right\}_{ac;bd}  \; .
\end{equation}
This implies that $|\psi\rangle_{abcd}$ is also a superposition
of double Bell states (albeit with different amplitudes) for
these two other partitions, which, therefore, also yield mixtures
of Bell states when tracing over half of the system.
Table~\ref{table_psi} summarizes the amplitudes of
$|\psi\rangle_{abcd}$ for the three partitions of $abcd$
into two pairs, starting from Eq.~(\ref{eq_psi}).
Thus, identifying $a$ with the reference qubit (initially 
entangled with the input), $b$ and $c$ with the two outputs, and
$d$ with an idle qubit (or the cloning machine),
Table~\ref{table_psi} defines the desired family of asymmetric Pauli
cloning machines.
\par

\begin{figure}
\caption{Ellipse delimiting the best quality of the two outputs of
an asymmetric PCM that can be achieved simultaneously
(only the quadrant $x,x'\ge 0$ is of interest here). The outputs
emerge from depolarizing channels of probability $p=3x^2$ 
and $p'=3x'^2$. Any close-to-perfect cloning characterized 
by a point inside the ellipse is forbidden.}
\vskip 0.25cm
\centerline{\psfig{file=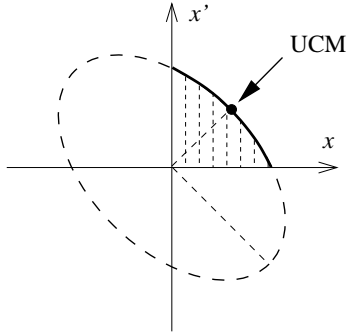,width=1.8in,angle=-90}}
\label{fig_nocloning}
\vskip -0.25cm
\end{figure}

Let us consider a PCM whose first output $b$ emerges from a depolarizing
channel of probability $p=3|x|^2$, i.e.,
\begin{equation}  \label{eq_asymm1}
\rho_{ab}=
 |v|^2 |\Phi^+\rangle\langle \Phi^+|
+|x|^2 \left( |\Phi^-\rangle\langle \Phi^-|
+|\Psi^+\rangle\langle \Psi^+|
+|\Psi^-\rangle\langle \Psi^-|  \right)  \; .
\end{equation}
From Table~\ref{table_psi},
the second output $c$ necessarily emerges from a depolarizing channel
of probability $p'={3\over 4}|v-x|^2$, or, more precisely,
\begin{equation}   \label{eq_asymm2}
\rho_{ac}= {|v+3x|^2 \over 4}
|\Phi^+\rangle\langle \Phi^+|
+{|v-x|^2 \over 4} \left( |\Phi^-\rangle\langle \Phi^-|
+ |\Psi^+\rangle\langle \Psi^+|
+ |\Psi^-\rangle\langle \Psi^-| \right)  \; .
\end{equation}
(The idle qubit $d$ emerges in general from a different Pauli channel.)
Clearly, both outputs of this asymmetric PCM are state-independent 
as they simply correspond to a (different) shrinking 
of the vector characterizing the input qubit in the Bloch sphere.
Combining Eqs.~(\ref{eq_asymm1}) and (\ref{eq_asymm2}),
the tradeoff between the quality of the two copies
can be described by the {\em no-cloning inequality}
\begin{equation}
x^2+x'^2+xx' \ge {1\over 4} \; ,   \qquad {\rm (with~}x,x'\ge 0)
\label{eq_no-cloning-ineq}
\end{equation}
where the copying error is measured by the probability of the
depolarizing channel underlying each output, i.e., 
$p=3x^2$ and $p'=3x'^2$.
(I assume here that that the amplitudes 
of $|\psi\rangle_{abcd}$ are real and positive.)
Equation~(\ref{eq_no-cloning-ineq}) corresponds to the domain
in the $(x,x')$-space located outside an ellipse
whose semiminor axis $b=1/\sqrt{6}$ is oriented 
in the direction $(1,1)$, as shown in Fig.~\ref{fig_nocloning}.
(The semimajor axis is $a=1/\sqrt{2}$.) 
The origin in this space corresponds to a (nonexisting) cloner whose
two outputs would be perfect $p=p'=0$. The ellipse characterizes
the ensemble of values for $p$ and $p'$ that can actually be achieved
with a PCM. It intercepts its minor axis
at $(1/\sqrt{12},1/\sqrt{12})$, which corresponds to the
universal cloning machine (UCM), i.e., $p=p'=1/4$. 
This point is the closest to the origin 
(i.e., the cloner with minimum $p+p'$), 
and characterizes in this sense the best possible copying, as
expected. Note that the underlying wave function
\begin{eqnarray}  \label{eq_psi-UCM}
|\psi\rangle_{abcd}
&=&\sqrt{3\over 4} \; |\Phi^+\rangle_{ab} |\Phi^+\rangle_{cd}
+\sqrt{1\over12} \left\{
|\Phi^-\rangle|\Phi^-\rangle
+|\Psi^+\rangle|\Psi^+\rangle
+|\Psi^-\rangle|\Psi^-\rangle \right\}_{ab;cd}  \\
&=& \sqrt{1\over 3} \left\{
|\Phi^+\rangle |\Phi^+\rangle
+|\Phi^-\rangle|\Phi^-\rangle
+|\Psi^+\rangle|\Psi^+\rangle \right\}_{ad;bc}   \; .
\end{eqnarray}
is symmetric under the interchange of $a$ and $d$ (or $b$ and $c$).
The ellipse crosses the $x$-axis at $(1/2,0)$, which describes the
situation where the first output emerges from a 100\%-depolarizing 
channel ($p=3/4$) while the second emerges from a perfect
channel ($p'=0$). Of course, $(0,1/2)$ corresponds to the symmetric
situation. The argument used in~\cite{fn0} strongly suggests that
the imperfect cloning achieved by such an asymmetric PCM is optimal:
a single additional qubit $d$ is sufficient to perform optimal
cloning, i.e., to achieve the minimum $p$ and $p'$ for a fixed ratio $p/p'$.
The domain inside the ellipse corresponds then to the values for $p$ and $p'$
that cannot be achieved simultaneously, reflecting the
impossibility of close-to-perfect cloning, and
Eq.~(\ref{eq_no-cloning-ineq})
is the tightest no-cloning bound that can be written for a qubit.
\par

\begin{figure}
\caption{Oblate ellipsoid representing the class of symmetric PCMs
whose two outputs emerge from the same Pauli channel
of parameters $p_x=x^2$, $p_y=y^2$, and $p_z=z^2$
(only the octant $x,y,z\ge 0$ is considered here). The pole of this
ellipsoid corresponds to the UCM.
The capacity of a Pauli channel that lies outside this ellipsoid
must be vanishing.}
\vskip 0.25cm
\centerline{\psfig{file=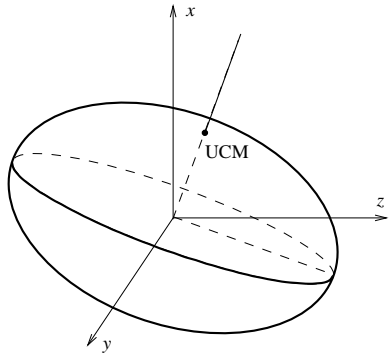,width=2.0in,angle=-90}}
\label{fig_ellipsoid}
\vskip -0.25cm
\end{figure}

Consider now the class of symmetric PCMs that have both outputs
emerging from a {\em same} Pauli channel, i.e., $\rho_{ab}=\rho_{ac}$.
Using Table~\ref{table_psi}, we obtain the conditions
$4|v|^2=|v+z+x+y|^2$, $4|z|^2=|v+z-x-y|^2$,
$4|x|^2=|v-z+x-y|^2$, and $4|y|^2=|v-z-x+y|^2$,
which yields
\begin{equation}  \label{eq_condition-dupl}
v=x+y+z  \; ,
\end{equation}
where $x$, $y$, $z$, and $v$ are assumed to be real.
Equation~(\ref{eq_condition-dupl}), together with the normalization
condition, describes a 2D surface in a space
where each point $(x,y,z)$ represents a Pauli channel of parameters
$p_x=x^2$, $p_y=y^2$, and $p_z=z^2$ (I only consider here the first
octant $x,y,z\ge 0$). This surface,
\begin{equation}  \label{eq_ellipsoid}
x^2+y^2+z^2+xy+xz+yz={1\over 2}  \; ,
\end{equation}
is an oblate ellipsoid $E$ with symmetry axis
along the direction $(1,1,1)$, as shown in Fig.~\ref{fig_ellipsoid}.
The semiminor axis (or polar radius)
is $a=1/2$ while the semimajor axis (or equatorial radius)
is $b=1$. In this representation, the distance to the origin 
is $p_x+p_y+p_z$,
so that the pole $(1/\sqrt{12},1/\sqrt{12},1/\sqrt{12})$
of this ellipsoid---the closest point to the origin---corresponds
to the special case of a depolarizing channel
of probability $p=1/4$. Thus, this particular
PCM coincides with the UCM. This simply illustrates that
the requirement of having an optimal cloning (minimum $p_x+p_y+p_z$)
implies that the cloner is state-independent ($p_x=p_y=p_z$).
\par

The class of symmetric PCMs 
characterized by Eq.~(\ref{eq_ellipsoid})
can be used in order to put a limit on the quantum capacity 
of a Pauli channel, thereby extending the result 
of Bruss et al.~\cite{bib_bruss}.
Indeed, applying an error-correcting scheme separately on each output
of the cloning machine (obliviously of the other output)
would lead to a violation of the no-cloning theorem if
the capacity $C$ was nonzero. Since $C$ is a nonincreasing
function of $p_x$, $p_y$, and $p_z$, for $p_x,p_y,p_z\le 1/2$
(i.e., adding noise to a channel cannot increase its capacity), 
I conclude that $C(p_x,p_y,p_z)=0$ for $(x,y,z)\not\in E$,
that is, the quantum capacity is vanishing for any Pauli channel
that lies outside the ellipsoid $E$. In particular,
Eq.~(\ref{eq_ellipsoid}) implies that the quantum capacity vanishes for
(i)~a depolarizing channel with $p=1/4$ ($p_x=p_y=p_z=1/12$)~\cite{bib_bruss}; 
(ii)~a ``2-Pauli'' channel with $p=1/3$ ($p_x=p_z=1/6$, $p_y=0$);
and (iii)~a dephasing channel with $p=1/2$ ($p_x=p_y=0$, $p_z=1/2$).
Furthermore, using the fact that $C$ cannot be
superadditive for a convex combination of a perfect 
and a noisy channel~\cite{bib_bdsw}, an upper bound on $C$
can be written using a linear interpolation between the perfect channel
$(0,0,0)$ and any Pauli channel lying on $E$:
\begin{equation}
C \le 1-2(x^2+y^2+z^2+xy+xz+yz) \; .
\end{equation}
Note that another class of symmetric PCMs can be found by
requiring $\rho_{ab}=\rho_{ad}$, which implies $v=x-y+z$
rather than Eq.~(\ref{eq_condition-dupl}). This requirement gives rise
to the reflection of $E$ with respect to the $xz$-plane, i.e, $y\to -y$. 
It does not change the above bound on $C$ because this class of PCMs
has noisier outputs in the octant $x,y,z\ge 0$.
\par

Let us now turn to the fully symmetric PCMs that have {\em three} outputs
emerging from the {\em same} Pauli channel, which corresponds
to a family of (non-optimal) quantum {\em triplicating} machines. 
The requirement $\rho_{ab}=\rho_{ac}=\rho_{ad}$
implies $(v=x+z) \wedge (y=0)$.
Incidentally, we notice
that if {\em all} pairs are required to be in the {\em same}
mixture of Bell states, this mixture cannot have
a singlet $|\Psi^-\rangle$ component. The outputs of the corresponding 
triplicators emerge therefore from a ``2-Pauli'' channel ($p_y=0$),
so that these triplicators are {\em state-dependent}, in contrast with
the one considered in Ref.~\cite{bib_gisinmassar}. (For describing
a {\em state-independent} triplicator, a 6-qubit wave function should be
used.) These triplicators are represented by the
intersection of $E$ with the $xz$-plane, that is, the ellipse
\begin{equation}
x^2+z^2+xz={1\over 2}  \; ,
\end{equation}
with semiminor axis $b=1/\sqrt{3}$ [oriented along the direction
$(1,1)$] and semimajor axis $a=1$. The intersection of this ellipse
with its semiminor axis ($x=z=1/\sqrt{6}$) corresponds to
the 4-qubit wave function
\begin{equation}  \label{eq_psi-triplicator}
|\psi\rangle_{abcd}={2\over\sqrt{6}}|\Phi^+\rangle|\Phi^+\rangle
+{1\over\sqrt{6}}|\Phi^-\rangle|\Phi^-\rangle
+{1\over\sqrt{6}}|\Psi^+\rangle|\Psi^+\rangle  \; ,
\end{equation}
which is symmetric under the interchange of any two qubits
and maximizes the 2-bit entropy (or minimizes the mutual entropy between
any two outputs of the triplicator). Equation~(\ref{eq_psi-triplicator}) thus
characterizes the best triplicator of this ensemble,
whose three outputs emerge from
a ``2-Pauli'' channel with $p=1/3$ ($p_x=p_z=1/6$). 
Equivalently, the (state-dependent)
operation of this triplicator on an arbitrary qubit can be written as
\begin{equation} \label{eq_best-triplicator}
|\psi\rangle \to {1\over 2}|\psi\rangle\langle\psi|
+ {1\over 6}|\psi^*\rangle\langle\psi^*|
+ {1 \over 3} (\openone /2)  \; .
\end{equation}
If $|\psi\rangle$ is real, Eq.~(\ref{eq_best-triplicator})
reduces to the triplicator 
that is considered in Ref.~\cite{bib_bh_tripl}. The fidelity of
cloning is then the same as for the UCM.
\par

I have generalized the Buzek-Hillery cloning machine and
introduced an asymmetric Pauli cloning machine, whose outputs emerge from
distinct Pauli channels. This allowed me to derive a tight 
no-cloning inequality for quantum bits, quantifying the impossibility
of copying due to quantum mechanics.
Furthermore, I have established an upper bound on the quantum
capacity of the Pauli channel relying on a class of symmetric PCMs.

\newpage
\acknowledgements

This work was supported in part by the National Science Foundation
under Grant Nos. PHY 94-12818 and PHY 94-20470,
and by a grant from DARPA/ARO through the QUIC Program
(\#DAAH04-96-1-3086).

\begin{table}
\begin{tabular}{c c c c c}
$|\psi\rangle_{abcd}$ &  $|\Phi^+\rangle |\Phi^+\rangle$ 
& $|\Phi^-\rangle |\Phi^-\rangle$ 
& $|\Psi^+\rangle |\Psi^+\rangle$ 
& $|\Psi^-\rangle |\Psi^-\rangle$ \\[1pt]
\hline
$ab$ vs $cd$ & $v$ & $z$ & $x$ & $y$ \\[1 pt]
$ac$ vs $bd$ & ${1\over 2}(v+z+x+y)$ & ${1\over 2}(v+z-x-y)$ &
          ${1\over 2}(v-z+x-y)$ & ${1\over 2}(v-z-x+y)$ \\[1 pt]
$ad$ vs $bc$ & ${1\over 2}(v+z+x-y)$ & ${1\over 2}(v+z-x+y)$ &
          ${1\over 2}(v-z+x+y)$ & ${1\over 2}(v-z-x-y)$
\end{tabular}
\caption{Amplitudes of $|\psi\rangle_{abcd}$
in terms of the double Bell states for the three possible partitions
of the four qubits $abcd$ into two pairs. }
\label{table_psi}
\end{table}


\begin{references}
\bibitem{fn0} It is shown in Refs.~\cite{bib_bh,bib_bruss} 
that a 2-dimensional ancilla space
is sufficient, so we only have to consider a single
qubit $y_3$ for the cloner.
\bibitem{fn1} Permutations of the Bell states in Eq.~(\ref{eq_psi})
are not considered for simplicity.
\end{references}
\end{document}